\documentclass[showpacs,amsmath,amssymb,aps,10pt,preprint,superscriptaddress]{revtex4-1}
\usepackage{graphicx}
\usepackage{bm}
\usepackage[breaklinks=true,colorlinks=true,linkcolor=blue,urlcolor=blue,citecolor=blue]{hyperref}
\usepackage{graphics}
\usepackage{dcolumn}
\usepackage{amsmath,amssymb}
\usepackage{mathdots}
\usepackage{natbib}
\usepackage{soul,color}
\usepackage{epstopdf}

\begin{document}

\title{Nonlinear relaxation between magnons and phonons in insulating ferromagnets}

\author{Valerij~A.~Shklovskij}
\affiliation{Physics Department, V. Karazin Kharkiv National University, 61077 Kharkiv, Ukraine}
\author{Viktoriia V. Mezinova}
\affiliation{Physics Department, V. Karazin Kharkiv National University, 61077 Kharkiv, Ukraine}
\author{Oleksandr~V.~Dobrovolskiy}
\email[Corresponding author: ]{Dobrovolskiy@Physik.uni-frankfurt.de}
\affiliation{Physikalisches Institut, Goethe University, 60438 Frankfurt am Main, Germany}
\affiliation{Physics Department, V. Karazin Kharkiv National University, 61077 Kharkiv, Ukraine}

\date{\today}

\begin{abstract}
Nonlinear relaxation between spin waves (magnons) and the crystal lattice (phonons) in an insulating ferromagnet is investigated theoretically. Magnons and phonons are described by the equilibrium Bose-Einstein distributions with different temperatures. The nonlinear heat current from magnons to phonons is calculated microscopically in terms of the Cherenkov radiation of phonons by magnons. The results are discussed in comparison with the well-known theoretical results on the nonlinear electron-phonon relaxation in metals  [Kaganov, Lifshitz, Tanatarov, \href{http://www.jetp.ac.ru/cgi-bin/e/index/e/4/2/p173?a=list}{J. Exp. Theor. Phys. \textbf{31}, 232 (1956)}]. The elaborated theoretical description is relevant for spin-pumping experiments and thermoelectric devices in which the magnon temperature is essentially higher than the phonon one.
\end{abstract}

\maketitle


\section{Introduction}

In the last years, spin caloritronics, which is concerned with the interplay between spin and heat currents in magnetic materials, has attracted great attention \cite{Bau12nam,Sch13prb,Boo14ees}. This attention is, in particular, motivated by recent discoveries related to thermal spin injection via the spin Seebeck effect \cite{Uch08nat,Wei13prl} that can produce spin current densities that are two orders of magnitude larger than those produced via electronic or resonant excitation approaches. For instance, within the context of energy conversion applications, thermal spin transport provides conceptually new mechanisms for solid-state thermal-to-electrical energy conversion that may be used for waste heat recovery and temperature control \cite{Boo14ees}. Furthermore, the field of magnon spintronics has emerged \cite{Chu15nph}, concerned with structures, devices and circuits that use spin currents carried by magnons, the quanta of spin waves. Analogous to conventional electric currents, magnon-based currents can be used to carry, transport and process information as alternative to charge-current-driven spintronic devices \cite{Chu14nac,Gru16nan}. Recently, pure magnonic spin currents in insulating ferromagnets featuring absence of Joule heating and reduced spin wave damping have been suggested for the implementation of efficient logic devices \cite{Cra18nac}. At the same time, spin waves can transport heat in the same manner as the lattice excitations (phonons) transport heat through perturbations of the atom positions \cite{San77prb,Ant13nam}. Heat transport by magnons and their relaxation on phonons become especially important in such insulating magnetic materials as, e.g. Y$_3$Fe$_5$O$_{12}$ \cite{Ser10jpd}, in contradistinction to metallic ferromagnets whose thermal conductivity is dominated by the conduction electrons.

While the electron-phonon and magnon-phonon relaxation has been investigated in a series of theoretical works \cite{Bha66prv,San77prb,Kag56etp,Kab08prb,Xia10prb,Bez16ltp,Liu17prb,Bez18pcm}, the \emph{nonlinear relaxation} of magnons on phonons --- the subject of this work --- has not been addressed theoretically so far. In this regard, the most closely related available theoretical work, which is similar in both, the problem statement and the solution scheme, is the problem of nonlinear relaxation of electrons on phonons considered by Kaganov, Lifschitz and Tanatarov (KLT) back in 1956 \cite{Kag56etp}. In that work, which is still the main model for analyzing experiments on the energy relaxation of excited electrons in metals \cite{Ani70boo,Ani74spj,Fuj84prl,Els87prl,Sch87prl}, the nonlinear heat current $Q$ from hot electrons at temperature $T_e$ to cold phonons at temperature $T_p$ in metals was calculated within the framework of the \emph{two-temperature model} with $T_e$ and $T_p$ being smaller than the Debye temperature $\Theta_D$. A nonlinear expression was obtained for the heat current $Q=A(T_e^5-T_p^5)$ from electrons to phonons, where $A$ is a constant expressed via the conductivity and the lattice parameters of the metal \cite{Kag56etp}. While the KLT results have allowed for analyzing various aspects of the time-dependent dynamics of hot electrons in metallic thin films at low temperatures $(T\ll\Theta_D)$ \cite{Bez16ltp,Bez18pcm}, so far the problem of relaxation between magnons and phonons in insulating ferromagnets has only been considered \cite{Akh46jpu} in the \emph{linear} regime $Q\sim(T_s-T_l)$, where $T_s$ is the magnon temperature. In state-of-the-art spin-pumping experiments \cite{Bau12nam,Sch13prb,Boo14ees,Jun15prb}, however, the magnon temperature $T_s$ can be essentially higher than the phonon temperature $T_l$, thus requiring a theoretical account for the nonlinear heat current regime.

Here, we bridge this gap by considering the case of \emph{nonlinear} relaxation between magnons and phonons when $T_s > T_l$ and derive expressions for the nonlinear heat current from magnons to phonons in an insulating ferromagnet.

\section{Main results}

\begin{figure}[t!]
    \centering
    \includegraphics[width=0.25\linewidth]{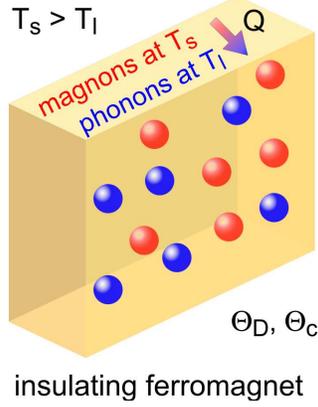}
    \caption{Formulation of the problem: The nonlinear relaxation between magnons and phonons (denoted as red and blue balls, respectively) is considered in an insulating ferromagnet ($\Theta_D$: Debye temperature, $\Theta_c$: Curie temperature). Magnons are characterized by the temperature $T_s$ which is essentially higher than the phonon temperature $T_l$, but much smaller than the Debye temperature $\Theta_D$. The magnon subsystem is considered in the quasi-equilibrium regime. The searched-for quantity is the nonlinear heat current $Q$ from hot magnons to cold phonons.}
    \label{f1}
\end{figure}

Specifically, we consider the following problem. The nonlinear relaxation between spin waves (magnons) and the crystal lattice (phonons) is considered in an insulating ferromagnet, Fig. \ref{f1}. In the ferromagnet, magnons are characterized by the temperature $T_s$ which is essentially higher than the phonon temperature $T_l$, i.e. $T_s> T_l$. The equilibration time for magnons in the ferromagnet is much smaller than the equilibration time between magnons and the crystal lattice \cite{Akh46jpu,Akh67boo}. Therefore, the magnon subsystem is considered in the quasi-equilibrium regime described by the conventional Bose-Einstein distribution $n(\varepsilon_k/T_s)=[\exp(\varepsilon_k/T_s)-1]^{-1}$, where $\varepsilon_k=\Theta_c(ak)^2$ is the dispersion law for magnons in the long-wavelength limit $ka\ll1$ with $\Theta_c$ being the Curie temperature of the ferromagnet, $a$ the lattice constant, and $k = |\mathbf{k}|$ the magnon wavevector. The theoretical task is to derive \emph{microscopically} the nonlinear heat current $Q$ from hot magnons at the temperature $T_s$ to cold phonons at the temperature $T_l$.

To accomplish this, we calculate the change in the number of phonons with the given wavevector $\mathbf{q}$ per unit of time $(\dot N_{\bf q})_s$ via the phonon-magnon collision integral $L_{ls} \lbrace N,n\rbrace$ \cite{Akh67boo} describing the absorption or emission of phonons by magnons, $(\dot N_{\bf q})_s = L_{ls} \lbrace N,n\rbrace$. Given the momentum conservation, $L_{ls} \lbrace N,n\rbrace$ can be expressed as
\begin{equation}
    \begin{array}{lll}
        \label{eColInt}
        L_{ls} \lbrace N,n\rbrace =\\[2mm]
        \hspace{0mm}\frac{2\pi}{\hbar}\sum_{\bf k}{|\psi_{sl}(\bf q,\bf k|\bf k +\bf q)|}^2 \lbrace(N_{\bf q} + 1)(n_{\bf k} +1) n_{\bf k +\bf q} -\\[2mm]
        \hspace{13mm}N_{\bf q}n_{\bf k}(n_{\bf k +\bf q} + 1)\rbrace\times\delta(\hbar\omega_q+\varepsilon_{\bf k } -\varepsilon_{\bf k +\bf q}).
    \end{array}
\end{equation}
Here, $|\psi_{sl}({\bf q},{\bf k}|{\bf k} +{\bf q})|^2$ is the squared matrix element of the transition probability. It reads \cite{Akh67boo}
\begin{equation}
    \label{eMatrixElement}
    |\psi_{sl}({\bf q},{\bf k}|{\bf k} +{\bf q})|^2 = \frac{{\Theta_c}^2}{N}(\frac{\hbar}{\rho a^3 \omega_q })a^4 k^2 {(\bf k +\bf q )}^2 q^2,
\end{equation}
where $\rho = M/{a^3}$, $M$ is the mass of the magnetic ion, $a$ is the lattice constant, $\Theta_c$ is the Curie temperature, $N$ is the number of atoms, $\omega_q = s q$ is the frequency of phonons with the wavevector $\bf q$, $s$ is the average speed of sound, and $\delta$ is the Dirac delta function.

In Eq. \eqref{eColInt}, $N_{\bf q}$ and $n_{\bf k}$ are the equilibrium Bose-Einstein distributions for phonons at the temperature $T_l$ and magnons at the temperature  $T_s$, namely
\begin{equation}
    \label{eBoseFunctions}
        N_{\bf q}=\frac{1}{\exp[({\hbar\omega_q}/{T_l})-1]},\hspace{5mm}
        n_{\bf k}=\frac{1}{\exp[({\varepsilon_k}/{T_s})-1]},
\end{equation}
where $\varepsilon_k=\Theta_c{(ak)}^2$ is the dispersion law for magnons in the long-wavelength limit $ka\ll 1$. In the limiting case $T_l=T_s$, from Eq. \eqref{eColInt} follows $L_{ls}{\lbrace N, n\rbrace}=0$.

With the calculation steps detailed in Appendix, the searched-for change in the number of phonons reads
\begin{equation}
    \label{eNdot}
    \begin{array}{lll}
    \dot{N}_{\bf q}=D(T_s)[n(\varepsilon_q/T_s )-n(\varepsilon_q/T_l )]
    \sum_{p=1}^\infty (1-e^{-px})\int_{y_0}^\infty dy (yx+y^2 ) e^{-py}.
    \end{array}
 \end{equation}
Here, $D(T_s)=(\Theta_c \Theta_D/8\pi\hbar\Theta_p)(T_s/\Theta_c)^3$, $\Theta_D=\hbar s/a$, $\Theta_p=Ms^2$, $x\equiv\varepsilon_q/T_l=\hbar\omega_q/T_l$, and $y_0=\Theta_D^2/4T\Theta_c$, which plays the role of an effective inverse temperature. In the integral over the dimensionless magnon energy $y=\varepsilon_k/T_s$, the lower integration limit $y_0$ reflects the Cherenkov character of the emission of phonons by magnons. Namely, \emph{only magnons whose energy is larger than $\Theta_D^2/4\Theta_c$ can emit phonons}.

With the passage from summation over $\mathbf{k}$ to integration and after the introduction of the magnon ``overheating'' parameter $\gamma=T_s/T_l$, the heat current $Q=\sum_{\bf q}(\hbar\omega_{\bf q}) \dot{N}_{\bf q}$ from magnons to phonons acquires the form
\begin{equation}
    \label{eQlong}
    \begin{array}{lll}
    Q=({N}/{8\pi^3})({\Theta_D^2 \Theta_c}/{2\hbar\Theta_p})({T_s}/{\Theta_c})^3\times
    [({T_s}/{\Theta_D})^4-({T_l}/{\Theta_D})^4] K(p),
    \end{array}
\end{equation}
where
\begin{equation}
    \label{eK}
    K(p)=\int_0^\infty \frac{u^3 du}{e^u-1} [J_D(T_s,x=u,y_0)-J_D(T_s,x=u/\gamma,y_0)]
\end{equation}
and
\begin{equation}
    \label{eJDT}
    J_D(T)=\sum_{p=1}^\infty (1-e^{-px})e^{-py_0}[x(\frac{y_0}{p}+\frac{1}{p^2})+(\frac{y_0^2}{p}+\frac{2y_0}{p^2}+\frac{2}{p^3})].
\end{equation}

\begin{figure}[t!]
    \centering
    \includegraphics[width=0.6\linewidth]{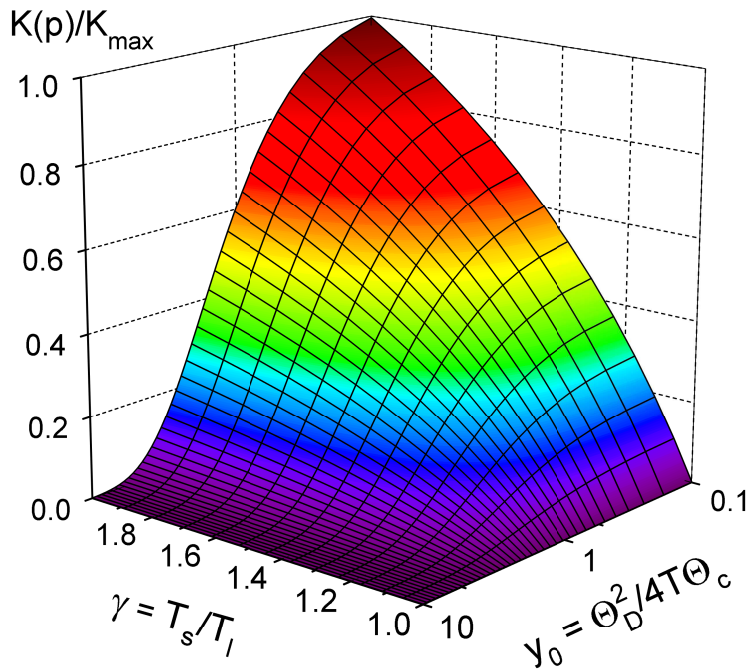}
    \caption{The integral $K(p)$ calculated by Eq. \eqref{eK} as a function of the magnon ``overheating'' parameter $\gamma = T_s/T_l$ and the effective inverse temperature $y_0 = \Theta^2_D / 4T\Theta_c$, normalized to its value $K_{max}$ at $\gamma = 2$ and $y_0 = 0.1$.}
    \label{f2}
\end{figure}

The dependence of the integral $K(p)$ on the parameter $\gamma = T_s/T_l$ and the effective inverse temperature $y_0 = \Theta^2_D / 4T\Theta_c$ is illustrated in Fig.~\ref{f2}. One sees that when the magnon and phonon temperatures are equal, i.e. when $\gamma = 1$, $K(p) = 0$ as expected. In the limiting case of large $y_0$, that corresponds to the limit of low temperatures, $K(p)$ becomes exponentially small due to the factor $\sim e^{-y_0}$  in Eq.~\eqref{eJDT}. The value of $K(p)$ increases with increase of both, the magnon ``overheating'' parameter $\gamma$ and the inverse temperature $y_0$.

While Eqs. \eqref{eQlong}--\eqref{eJDT} are valid at any arbitrary temperature $T_l$ when $T_s \ll \Theta_D$, the condition $T_s \ll \Theta_D$ allows us to essentially simplify Eq. \eqref{eJDT} in the low-temperature limit. Namely, we can limit ourselves to $p=1$ when $y_0(T_s)=\Theta_D^2/4T_s\Theta_c \gg 1$, since $J_D(T_s)\sim e^{-2y_0 }\ll 1$ for $p=2$. Namely, at $T_s \ll \Theta_D$
\begin{equation}
    \label{eK1}
    \begin{array}{lll}
    K(p=1)=\varphi_1 \Gamma(5)[1+\mu[\zeta(5,1+\mu)-\zeta(5)]]+
    \varphi_2\Gamma(4)[1+\mu[\zeta(4,1+\mu)-\zeta(4)]],
    \end{array}
\end{equation}
where $\Gamma(n)$ is the gamma function, $\zeta(n,m)$ is the generalized zeta function, $\mu=1/\gamma=T_l/T_s$, $ \varphi_1=e^{-y_0 }(y_0+1)$ and $\varphi_2=e^{-y_0 }(y_0^2+2y_0+2)$. The final result for $Q(p=1)$ is obtained by substituting Eq. \eqref{eK1} into Eq. \eqref{eQlong}.

\section{Discussion}

Proceeding to a discussion of the obtained results, first of all we recall that Eqs. \eqref{eQlong}--\eqref{eJDT} describe the nonlinear heat current between \emph{magnons and phonons} in an insulating ferromagnet in the case when the states of the magnon and phonon subsystems are described by the equilibrium Bose-Einstein distributions with different temperatures $T_s$ and $T_l$, respectively. Experimentally, the condition $T_s>T_l$ can be realized in consequence of, e.g., parametric pumping of spin waves in insulating ferromagnets \cite{Akh67boo}. Theoretically, the formulation of the considered problem is conceptually similar to the two-temperature KLT problem \cite{Kag56etp} of nonlinear relaxation between \emph{electrons and phonons} in a metallic sample. Since the KLT model is widely used for analyzing experiments on the energy relaxation of excited electrons in metals  \cite{Ani74spj,Fuj84prl,Els87prl,Sch87prl,Sch13prb}, in what follows it is instructive to briefly outline the main results of the KLT work with the aid of emphasizing its similarities and differences with the magnon-phonon nonlinear relaxation problem considered here.

Specifically, the KLT work relies upon a quadratic and isotropic dispersion of the electron energy in a metal $\epsilon_{\bf p} = p^2/2m$, where $m$ is the effective mass. It is assumed that phonons have only a longitudinal acoustic mode with the linear dispersion $\omega_{q} = sq$, where $s$ is the speed of longitudinal sound and $q = |\mathbf{q}|$ is the phonon wavevector. KLT use a deformation potential approximation for the electron-phonon interaction (EPI) \cite{Kag56etp}. Namely, the probability of the electron transition from the state with momentum ${\bf p}$ into the state with momentum ${\bf p}\prime$ per unit of time is expressed by the function $w(q)$ which is proportional to the squared EPI matrix element
\begin{equation}\label{2}
    w(q)={{\pi \mu^2  \omega_q}\over{\rho_f s^2}},
\end{equation}
where $\mu$ is the constant of the deformation potential on the order of the Fermi energy $\mu \sim \varepsilon_F=p_F^2/2m$ and $\rho_f$ is the film density. In the KLT work, electrons and phonons are considered in quasi-equilibrium and they are characterized by the temperatures $T_e$ and $T_p$, respectively.

For the derivation of the dynamic equations for the electron and phonon temperatures KLT derived the specific power $P_{ep}$ of the heat current from hot electrons to cold phonons, which is expressed via the electron-phonon collision integral
\begin{equation}
    \label{ePep}
    P_{ep}=\int {{d^3q}\over{(2\pi)^3}}\, \hbar \omega_q \,I_{pe}(N_{\bf q}, f_{\bf p}).
\end{equation}
With the Bose-Einstein distribution $N_{\bf q}=n_q \equiv [\exp(\hbar\omega_q/k_B\,T_p)-1]^{-1}$ for phonons and the Fermi distribution $f_{\bf p}=f_0(\epsilon_{p}) \equiv \{\exp[(\epsilon_{p}-\epsilon_F)/k_B\,T_e]+1\}^{-1}$ for electrons, KLT obtained the following expression for $P_{ep}$, which is valid at arbitrary temperatures \cite{Kag56etp}
\begin{equation}
    \label{ePepSolution}
    P_{ep}(T_e,T_p)={\frac {m^2 \mu^2 (k_B \Theta_D)^5}{4 \pi^3\hbar^7 \rho_f s^4}}[F(T_e)-F(T_p)],
\end{equation}
where the function $F(T)$ is determined by
\begin{equation}
    \label{eFvT}
    F(T)= \Bigl({{T}\over{\Theta_D}}\Bigr)^5\int_{0}^{\Theta_D/T}{\frac {x^4\,dx}{e^x-1}}.
\end{equation}
From Eqs. \eqref{ePepSolution} and \eqref{eFvT} it follows that at high temperatures (with respect to $\Theta_D$) $P_{ep}=\alpha(T_e-T_p)$, while at low temperatures $P_{ep}=A(T_{e}^5-T_{p}^5)$. The constants $\alpha=( {m^2 \mu^2 k_{B}^5 \Theta_{D}^4)/(16 \pi^3\hbar^7 \rho_f s^4})$ and $A=({D_5 m^2 \mu^2 k_{B}^5)/(4 \pi^3\hbar^7 \rho_{f} s^4})$ do not depend on the electron and phonon temperatures and determine the strength of the EPI at high and low temperatures, respectively. In the last equality, $D_5 \approx 24.9$ is the integration result of $D_k=\int_{0}^{\infty}x^{k-1}(e^x -1)^{-1}\, dx$ at $k=5$. On the basis of the KLT work \cite{Kag56etp} one can write down the system of the nonlinear dynamic equations for the electron and phonon temperatures \cite{Ani70boo,Ani74spj}. In the spatially homogenous case, which is typical for thin films, this system of equations reads
\begin{equation}
    \label{eCeTe}
        c_e(T_e){\frac {d\,T_e}{d\, t}}=-P_{ep}(T_e,T_p)+W(t),
\end{equation}
\begin{equation}
    \label{eCpTp}
        c_p(T_p){\frac {d\,T_p}{d\, t}}=P_{ep}(T_e,T_p),
\end{equation}
where $c_e$ and $c_p$ are the electron and magnon specific heats, respectively, and $W(t)$ is the specific power of heat sources heating the electrons.

Turning back to our magnon-phonon problem, in the spatially homogenous case of an insulating ferromagnetic thin film with $d<s/\nu_{ls}$, where $d$ is the film thickness and $\nu_{ls}$ is the collision frequency of phonons with magnons, we can write a system of the nonlinear dynamic equations for the magnon and phonon temperatures
\begin{equation}
    \label{eCsTs}
c_s(T_s){\frac {d\,T_s}{d\, t}}=-Q(T_s,T_l)+W_s(t),
\end{equation}
\begin{equation}
    \label{eClTl}
c_l(T_l){\frac {d\,T_l}{d\, t}}=Q(T_s,T_l),
\end{equation}
where $c_s$ and $c_l$ are the magnon and phonon specific heats, respectively, and $W_s(t)$ is the specific power of heat sources heating the magnons.

Now, we are in position to emphasize the similarities and the differences in the results obtained in the problems of nonlinear magnon-phonon relaxation in our work and the nonlinear electron-phonon relaxation in the KLT work.

Firstly, the general scheme for the calculation of the heat flows in both problems is formally similar, relying upon the formulae
\begin{equation}
    \label{eQmagnon}
    Q=\sum_{\bf q} \hbar\omega_{\bf q} \dot{N}_{\bf q}(T_s,T_l),
\end{equation}
\begin{equation}
    \label{eQelectron}
    P_{ep}=\sum_{\bf q} \hbar\omega_{\bf q} \dot{N}_{\bf q}(T_e,T_p),
\end{equation}
where $\dot{N}_{\bf q}$ is the change in the number of phonons with the wavevector $\bf q$ per unit of time. This change in the number of phonons is caused by the emission or absorption of phonons by magnons [Eq. \eqref{eQmagnon}] or electrons [Eq. \eqref{eQelectron}], and it is determined by the collision integrals \eqref{eColInt} and \eqref{ePep} for phonons with the respective quasiparticles. Both these collision integrals are equal to the product of the frequency $\nu$ of the collisions of phonons with magnons or electrons and the difference of the equilibrium Bose-Einstein distributions $n(\varepsilon_q/T)$, namely
\begin{equation}
    \label{eLls}
    L_{ls}=\nu_{ls}[n(\varepsilon_q/T_s)-n(\varepsilon_q/T_l)],
\end{equation}
\begin{equation}
    \label{eLpe}
    I_{pe}=\nu_{pe}[n(\varepsilon_q/T_e)-n(\varepsilon_q/T_p)],
\end{equation}
where $\varepsilon_q=\hbar\omega_q$ is the phonon energy.

Secondly, we note that while the integrals $L_{ls}$ and $I_{pe}$ in Eqs. \eqref{eLls} and \eqref{eLpe} look formally similar, the collision integral for magnons and phonons $L_{ls}$ given by Eq. \eqref{eColInt} for the collision frequency $\nu_{ls}$ has a more complex structure than that for the collision frequency of phonons with electrons  $\nu_{pe}\sim(s/\vartheta_F)\omega_q$ given by Eq. \eqref{ePep}.

Thirdly, the presence of the finite integration limit $y_0$ over the dimensionless magnon energy  $y=\varepsilon_k/T$ in Eq. \eqref{eNdot} is caused by the fact that the emission of phonons by magnons is only possible for magnons whose energy is larger than $\Theta_D^2/4\Theta_c$. It is this crucial point which underlines the \emph{Cherenkov character} of emission of phonons by magnons in insulating ferromagnets. This is distinct from the EPI in metals where any electron at the Fermi surface can absorb and emit a phonon, since the speed of sound in metals $s$ is much smaller than the electron Fermi velocity $\vartheta_F$. In consequence of this, in contrast to the frequency of the phonon-electron collisions $\nu_{pe}$ which only depends on the absolute value of the phonon wavevector $q$, the frequency of the phonon-magnon collisions in Eq. \eqref{eNdot} also \emph{depends on the magnon temperature} $T_s$, that is
\begin{equation}
    \label{eNuls}
    \nu_{ls}(T_s,q)=D(T_s)J_D(T_s).
\end{equation}

In addition, we note that the expression for $Q$ in Eq. \eqref{eQlong} is only valid when $T_s\ll\Theta_D$, while for electrons in metals the expression $P_{ep}(T_e,T_p)$ is valid at \emph{any arbitrary} $T_e$ и $T_p$ when $T_e\ll\varepsilon_F$. The same considerations hold for the nonlinear dynamic equations for electrons [Eqs. \eqref{eCeTe} and \eqref{eCpTp}] and magnons [Eqs. \eqref{eCsTs} and \eqref{eClTl}].

Finally, we would like to emphasize the general importance of the obtained results. In the experimental work by Schreier \emph{et al} \cite{Sch13prb} it has been pointed out that one of the challenges in analyzing the intertwinned charge, spin and heat currents in hybrid magnetic structures is a proper account for temperature differences in the electron, magnon and phonon subsystems, caused by the different thermal properties and boundary conditions for the respective quasiparticles. The phonon, electron, and magnon temperature profiles in substrate/ferromagnet/normal metal multilayers can exhibit discontinuities at the material interfaces due to interface properties such as the Kapitza resistance \cite{Kap41jph}. The temperature profiles are not easily measurable for a nonequilibrium situation in which magnon, phonon, and electron temperatures differ. An in depth analysis and interpretation of experimental spin Seebeck effect data is to date possible only by modeling the magnon, phonon, and electron temperature profiles based on the relevant material parameters \cite{Sch13prb}. Especially for magnetic insulators the determination of the phonon temperature $T_p$ profile is of central importance. Accordingly, the elaborated theoretical account for the nonlinear heat current from hot magnons and to cold phonons in insulating ferromagnets sets the foundation for a follow-up analysis of the magnon and phonon temperature profiles in multilayer spin caloritronic structures.

To conclude, we have theoretically investigated the nonlinear relaxation between magnons and phonons in an insulating ferromagnet. Magnons and phonons were described by the equilibrium Bose-Einstein distributions with different temperatures. The nonlinear heat current from magnons to phonons has been calculated microscopically in terms of the Cherenkov radiation of phonons by magnons. The elaborated theoretical account is relevant for spin-pumping experiments and thermoelectric devices in which the magnon temperature is essentially higher than the phonon one.

\begin{acknowledgments}
Research leading to this results received funding from the European Commission in the framework of the program Marie Sklodowska-Curie Actions --- Research and Innovation Staff Exchange (MSCA-RISE) under Grant Agreement No. 644348 (MagIC).
\end{acknowledgments}

\section*{Appendix}
This Appendix addresses the calculation of the collision integral given by Eq. \eqref{eColInt}. To this end, the curly bracket in Eq. \eqref{eColInt} is denoted by $\Phi$ and the new variables $x\equiv\varepsilon_q/T_l=\hbar\omega_q/T_l$ and $y\equiv\varepsilon_k/T_s$ are introduced. Then, $\Phi$ acquires the form
\begin{equation}
    \label{ePhi}
    \Phi=(\frac{1}{e^x-1}-\frac{1}{e^{x\gamma}-1})[\frac{e^y}{e^y-1}-\frac{e^{y+x}}{e^{y+x}-1}],
\end{equation}
where $\gamma=T_s/T_l >1$. Here, we have used the relations
\begin{eqnarray*}
    \label{eRelation1}
    \frac{1}{(e^{y+x}-1)(e^y-1)}= \frac{1}{e^x-1}[\frac{1}{e^y-1}-\frac{e^x}{e^{y+x}-1}],
\end{eqnarray*}
\begin{eqnarray*}
    \label{eRelation2}
    \frac{e^{\gamma x}-e^x}{(e^{\gamma x}-1)(e^x-1)}=\frac{1}{e^x-1}-\frac{1}{e^{\gamma x-1}}.
\end{eqnarray*}

The condition \eqref{ePhi} for $\Phi$ can be rewritten in terms of a sum of the geometric sequences with the decreasing denominators $e^{-y}$ and $e^{-(x+y)}$, namely
\begin{eqnarray*}
    \label{ePhicompact}
    \Phi= [n(\varepsilon_k/T_s )-n(\varepsilon_q/T_l )] \sum_{p=1}^\infty e^{-py} (1-e^{-px}).
\end{eqnarray*}

While passing from the sum over $\bf k$ to integration in Eq. \eqref{ePhi} in the long-wavelength limit $ka\ll 1$ we have used
\begin{equation*}
    \label{eSumToInt}
    \sum_{\bf{k}}\rightarrow\frac{V}{(2\pi)^3}\int d{\bf{k}}=\frac{N a^3}{(2\pi)^3}\int k^2 dk dO,
 \end{equation*}
where $dO=2\pi\sin\theta d\theta$ and $\theta$ is the polar angle of the vector $\bf k$ with respect to the vector $\bf{q}$. Given that
\begin{equation*}
    \label{eAngleRelation}
    \delta(\hbar\omega_q + \varepsilon_{\bf{k}} -\varepsilon_{{\bf{q}}+{\bf{k}}})= \frac{\delta(f-\cos\theta)}{\Theta_c (2a^2qk)},
\end{equation*}
where $f=({1}/{2ak})(({\Theta_D}/{\Theta_c}) - q a)$, one obtains Eq. \eqref{eNdot}:
\begin{equation*}
    \begin{array}{lll}
    \dot{N}_{\bf q}=D(T_s)[n(\varepsilon_q/T_s )-n(\varepsilon_q/T_l )]\times\sum_{p=1}^\infty (1-e^{-px})\int_{y_0}^\infty dy (yx+y^2 ) e^{-py},
    \end{array}
 \end{equation*}
where $D(T)=(\Theta_c \Theta_D/8\pi\hbar\Theta_p)(T/\Theta_c )^3$, $y_0=\Theta_D^2/4T\Theta_c$, $\Theta_D=\hbar s/a$, and $\Theta_p=Ms^2$. For the calculation of $$J_D (T)= \sum_{p=1}^\infty (1-e^{-px}) \int_{y_0}^\infty dy (yx+y) e^{-py}$$ one rewrites it as
\begin{equation}
    J_D(T)=\sum_{p=1}^\infty (1-e^{-px})e^{-py_0}[x(\frac{y_0}{p}+\frac{1}{p^2})+(\frac{y_0^2}{p}+\frac{2y_0}{p^2}+\frac{2}{p^3})].
\end{equation}
Noting that $J_D(T_s)\sim e^{-2y_0 }\ll 1$ for $p=2$ since $y_0(T_s)=\Theta_D^2/4T_s\Theta_c \gg 1$ we can limit ourselves by $p=1$, obtaining
\begin{equation*}
    J_D (T_s,p=1)\approx (1-e^{-x})e^{-y_0} [x(y_0+1)+y_0^2+2y_0+2].
\end{equation*}

The heat current from magnons to phonons is determined by
\begin{equation*}
    \begin{array}{lll}
    Q=\sum_{\bf q}(\hbar\omega_{\bf q}) \dot{N}_{\bf q}=
    \sum_{\bf q} (\hbar\omega_q)D(T_s)J_D(T_s,x,y_0)[n(\varepsilon_q/T_s)-n(\varepsilon_q/T_l)],
    \end{array}
\end{equation*}
where $J_D(T_s,x,y_0)$ is given by Eq. (23). By passing from $\sum_{\bf q}$ to the integral one can show that
\begin{equation}
    \label{24}
    \begin{array}{lll}
    Q=({N}/{8\pi^3})({\Theta_D^2 \Theta_c}/{2\hbar\Theta_p})({T_s}/{\Theta_c})^3\times
    [({T_s}/{\Theta_D})^4-({T_l}/{\Theta_D})^4]\times\\[3mm]
    \hspace{5mm}\int_0^\infty(\displaystyle\frac{u^3 du}{e^u-1})[J_D (T_s,x=u,y_0 )-J_D (T_s,x=u/\gamma,y_0)].
    \end{array}
\end{equation}
Here, the calculation of $Q$ at an arbitrary $p$ is reduced to the calculation of the integral
\begin{equation*}
    K(p)=\int_0^\infty \frac{u^3 du}{e^u-1} [J_D(T_s,x=u,y_0)-J_D(T_s,x=u/\gamma,y_0)].
\end{equation*}
Using relation 2.3.13.22 in Ref. \cite{Pru02boo} for $p=1$ one can rewrite
$$\int_0^\infty (u^{n-1} e^{-u} du)/(e^u-1)=\Gamma(n)[\zeta(n,2)],$$
where $\Gamma(n)$ is the gamma function and $\zeta(n,2)$ is the generalized zeta function. Then
\begin{equation}
    \label{25}
    \begin{array}{lll}
    K(p=1)=\varphi_1 \Gamma(5)[1+\mu[\zeta(5,1+\mu)-\zeta(5)]]+
    \varphi_2\Gamma(4)[1+\mu[\zeta(4,1+\mu)-\zeta(4)]].
    \end{array}
\end{equation}
Here, $ \varphi_1=e^{-y_0 }(y_0+1)$, $\varphi_2=e^{-y_0 }(y_0^2+2y_0+2)$, and $\mu=1/\gamma=T_l/T_s$. The final result for $Q(p=1)$ is obtained by the substitution of Eq. \eqref{25} into Eq. \eqref{24}.


\begin{thebibliography}{30}%
\makeatletter
\providecommand \@ifxundefined [1]{%
 \@ifx{#1\undefined}
}%
\providecommand \@ifnum [1]{%
 \ifnum #1\expandafter \@firstoftwo
 \else \expandafter \@secondoftwo
 \fi
}%
\providecommand \@ifx [1]{%
 \ifx #1\expandafter \@firstoftwo
 \else \expandafter \@secondoftwo
 \fi
}%
\providecommand \natexlab [1]{#1}%
\providecommand \enquote  [1]{``#1''}%
\providecommand \bibnamefont  [1]{#1}%
\providecommand \bibfnamefont [1]{#1}%
\providecommand \citenamefont [1]{#1}%
\providecommand \href@noop [0]{\@secondoftwo}%
\providecommand \href [0]{\begingroup \@sanitize@url \@href}%
\providecommand \@href[1]{\@@startlink{#1}\@@href}%
\providecommand \@@href[1]{\endgroup#1\@@endlink}%
\providecommand \@sanitize@url [0]{\catcode `\\12\catcode `\$12\catcode
  `\&12\catcode `\#12\catcode `\^12\catcode `\_12\catcode `\%12\relax}%
\providecommand \@@startlink[1]{}%
\providecommand \@@endlink[0]{}%
\providecommand \url  [0]{\begingroup\@sanitize@url \@url }%
\providecommand \@url [1]{\endgroup\@href {#1}{\urlprefix }}%
\providecommand \urlprefix  [0]{URL }%
\providecommand \Eprint [0]{\href }%
\providecommand \doibase [0]{http://dx.doi.org/}%
\providecommand \selectlanguage [0]{\@gobble}%
\providecommand \bibinfo  [0]{\@secondoftwo}%
\providecommand \bibfield  [0]{\@secondoftwo}%
\providecommand \translation [1]{[#1]}%
\providecommand \BibitemOpen [0]{}%
\providecommand \bibitemStop [0]{}%
\providecommand \bibitemNoStop [0]{.\EOS\space}%
\providecommand \EOS [0]{\spacefactor3000\relax}%
\providecommand \BibitemShut  [1]{\csname bibitem#1\endcsname}%
\let\auto@bib@innerbib\@empty
\bibitem [{\citenamefont {Bauer}\ \emph {et~al.}(2012)\citenamefont {Bauer},
  \citenamefont {Saitoh},\ and\ \citenamefont {van Wees}}]{Bau12nam}%
  \BibitemOpen
  \bibfield  {author} {\bibinfo {author} {\bibfnamefont {G.~E.~W.}\
  \bibnamefont {Bauer}}, \bibinfo {author} {\bibfnamefont {E.}~\bibnamefont
  {Saitoh}}, \ and\ \bibinfo {author} {\bibfnamefont {B.~J.}\ \bibnamefont {van
  Wees}},\ }\href {http://dx.doi.org/10.1038/nmat3301} {\bibfield  {journal}
  {\bibinfo  {journal} {Nat. Mater.}\ }\textbf {\bibinfo {volume} {11}},\
  \bibinfo {pages} {391 EP } (\bibinfo {year} {2012})}\BibitemShut {NoStop}%
\bibitem [{\citenamefont {Schreier}\ \emph {et~al.}(2013)\citenamefont
  {Schreier}, \citenamefont {Kamra}, \citenamefont {Weiler}, \citenamefont
  {Xiao}, \citenamefont {Bauer}, \citenamefont {Gross},\ and\ \citenamefont
  {Goennenwein}}]{Sch13prb}%
  \BibitemOpen
  \bibfield  {author} {\bibinfo {author} {\bibfnamefont {M.}~\bibnamefont
  {Schreier}}, \bibinfo {author} {\bibfnamefont {A.}~\bibnamefont {Kamra}},
  \bibinfo {author} {\bibfnamefont {M.}~\bibnamefont {Weiler}}, \bibinfo
  {author} {\bibfnamefont {J.}~\bibnamefont {Xiao}}, \bibinfo {author}
  {\bibfnamefont {G.~E.~W.}\ \bibnamefont {Bauer}}, \bibinfo {author}
  {\bibfnamefont {R.}~\bibnamefont {Gross}}, \ and\ \bibinfo {author}
  {\bibfnamefont {S.~T.~B.}\ \bibnamefont {Goennenwein}},\ }\href {\doibase
  10.1103/PhysRevB.88.094410} {\bibfield  {journal} {\bibinfo  {journal} {Phys.
  Rev. B}\ }\textbf {\bibinfo {volume} {88}},\ \bibinfo {pages} {094410}
  (\bibinfo {year} {2013})}\BibitemShut {NoStop}%
\bibitem [{\citenamefont {Boona}\ \emph {et~al.}(2014)\citenamefont {Boona},
  \citenamefont {Myers},\ and\ \citenamefont {Heremans}}]{Boo14ees}%
  \BibitemOpen
  \bibfield  {author} {\bibinfo {author} {\bibfnamefont {S.~R.}\ \bibnamefont
  {Boona}}, \bibinfo {author} {\bibfnamefont {R.~C.}\ \bibnamefont {Myers}}, \
  and\ \bibinfo {author} {\bibfnamefont {J.~P.}\ \bibnamefont {Heremans}},\
  }\href {\doibase 10.1039/C3EE43299H} {\bibfield  {journal} {\bibinfo
  {journal} {Energy Environ. Sci.}\ }\textbf {\bibinfo {volume} {7}},\ \bibinfo
  {pages} {885} (\bibinfo {year} {2014})}\BibitemShut {NoStop}%
\bibitem [{\citenamefont {Uchida}\ \emph {et~al.}(2008)\citenamefont {Uchida},
  \citenamefont {Takahashi}, \citenamefont {Harii}, \citenamefont {Ieda},
  \citenamefont {Koshibae}, \citenamefont {Ando}, \citenamefont {Maekawa},\
  and\ \citenamefont {Saitoh}}]{Uch08nat}%
  \BibitemOpen
  \bibfield  {author} {\bibinfo {author} {\bibfnamefont {K.}~\bibnamefont
  {Uchida}}, \bibinfo {author} {\bibfnamefont {S.}~\bibnamefont {Takahashi}},
  \bibinfo {author} {\bibfnamefont {K.}~\bibnamefont {Harii}}, \bibinfo
  {author} {\bibfnamefont {J.}~\bibnamefont {Ieda}}, \bibinfo {author}
  {\bibfnamefont {W.}~\bibnamefont {Koshibae}}, \bibinfo {author}
  {\bibfnamefont {K.}~\bibnamefont {Ando}}, \bibinfo {author} {\bibfnamefont
  {S.}~\bibnamefont {Maekawa}}, \ and\ \bibinfo {author} {\bibfnamefont
  {E.}~\bibnamefont {Saitoh}},\ }\href {http://dx.doi.org/10.1038/nature07321}
  {\bibfield  {journal} {\bibinfo  {journal} {Nature}\ }\textbf {\bibinfo
  {volume} {455}},\ \bibinfo {pages} {778 EP } (\bibinfo {year}
  {2008})}\BibitemShut {NoStop}%
\bibitem [{\citenamefont {Weiler}\ \emph {et~al.}(2013)\citenamefont {Weiler},
  \citenamefont {Althammer}, \citenamefont {Schreier}, \citenamefont {Lotze},
  \citenamefont {Pernpeintner}, \citenamefont {Meyer}, \citenamefont {Huebl},
  \citenamefont {Gross}, \citenamefont {Kamra}, \citenamefont {Xiao},
  \citenamefont {Chen}, \citenamefont {Jiao}, \citenamefont {Bauer},\ and\
  \citenamefont {Goennenwein}}]{Wei13prl}%
  \BibitemOpen
  \bibfield  {author} {\bibinfo {author} {\bibfnamefont {M.}~\bibnamefont
  {Weiler}}, \bibinfo {author} {\bibfnamefont {M.}~\bibnamefont {Althammer}},
  \bibinfo {author} {\bibfnamefont {M.}~\bibnamefont {Schreier}}, \bibinfo
  {author} {\bibfnamefont {J.}~\bibnamefont {Lotze}}, \bibinfo {author}
  {\bibfnamefont {M.}~\bibnamefont {Pernpeintner}}, \bibinfo {author}
  {\bibfnamefont {S.}~\bibnamefont {Meyer}}, \bibinfo {author} {\bibfnamefont
  {H.}~\bibnamefont {Huebl}}, \bibinfo {author} {\bibfnamefont
  {R.}~\bibnamefont {Gross}}, \bibinfo {author} {\bibfnamefont
  {A.}~\bibnamefont {Kamra}}, \bibinfo {author} {\bibfnamefont
  {J.}~\bibnamefont {Xiao}}, \bibinfo {author} {\bibfnamefont {Y.-T.}\
  \bibnamefont {Chen}}, \bibinfo {author} {\bibfnamefont {H.~J.}~\bibnamefont
  {Jiao}}, \bibinfo {author} {\bibfnamefont {G.~E.~W.}\ \bibnamefont {Bauer}},
  \ and\ \bibinfo {author} {\bibfnamefont {S.~T.~B.}\ \bibnamefont
  {Goennenwein}},\ }\href {\doibase 10.1103/PhysRevLett.111.176601} {\bibfield
  {journal} {\bibinfo  {journal} {Phys. Rev. Lett.}\ }\textbf {\bibinfo
  {volume} {111}},\ \bibinfo {pages} {176601} (\bibinfo {year}
  {2013})}\BibitemShut {NoStop}%
\bibitem [{\citenamefont {Chumak}\ \emph {et~al.}(2015)\citenamefont {Chumak},
  \citenamefont {Vasyuchka}, \citenamefont {Serga},\ and\ \citenamefont
  {Hillebrands}}]{Chu15nph}%
  \BibitemOpen
  \bibfield  {author} {\bibinfo {author} {\bibfnamefont {A.~V.}\ \bibnamefont
  {Chumak}}, \bibinfo {author} {\bibfnamefont {V.~I.}\ \bibnamefont
  {Vasyuchka}}, \bibinfo {author} {\bibfnamefont {A.~A.}\ \bibnamefont
  {Serga}}, \ and\ \bibinfo {author} {\bibfnamefont {B.}~\bibnamefont
  {Hillebrands}},\ }\href {http://dx.doi.org/10.1038/nphys3347} {\bibfield
  {journal} {\bibinfo  {journal} {Nat. Phys.}\ }\textbf {\bibinfo {volume}
  {11}},\ \bibinfo {pages} {453} (\bibinfo {year} {2015})}\BibitemShut
  {NoStop}%
\bibitem [{\citenamefont {Chumak}\ \emph {et~al.}(2014)\citenamefont {Chumak},
  \citenamefont {Serga},\ and\ \citenamefont {Hillebrands}}]{Chu14nac}%
  \BibitemOpen
  \bibfield  {author} {\bibinfo {author} {\bibfnamefont {A.~V.}\ \bibnamefont
  {Chumak}}, \bibinfo {author} {\bibfnamefont {A.~A.}\ \bibnamefont {Serga}}, \
  and\ \bibinfo {author} {\bibfnamefont {B.}~\bibnamefont {Hillebrands}},\
  }\href {http://dx.doi.org/10.1038/ncomms5700} {\bibfield  {journal} {\bibinfo
   {journal} {Nat. Commun.}\ }\textbf {\bibinfo {volume} {5}},\ \bibinfo
  {pages} {4700} (\bibinfo {year} {2014})}\BibitemShut {NoStop}%
\bibitem [{\citenamefont {Grundler}(2016)}]{Gru16nan}%
  \BibitemOpen
  \bibfield  {author} {\bibinfo {author} {\bibfnamefont {D.}~\bibnamefont
  {Grundler}},\ }\href {http://dx.doi.org/10.1038/nnano.2016.16} {\bibfield
  {journal} {\bibinfo  {journal} {Nat. Nanotech.}\ }\textbf {\bibinfo {volume}
  {11}},\ \bibinfo {pages} {407} (\bibinfo {year} {2016})}\BibitemShut
  {NoStop}%
\bibitem [{\citenamefont {Cramer}\ \emph {et~al.}(2018)\citenamefont {Cramer},
  \citenamefont {Fuhrmann}, \citenamefont {Ritzmann}, \citenamefont {Gall},
  \citenamefont {Niizeki}, \citenamefont {Ramos}, \citenamefont {Qiu},
  \citenamefont {Hou}, \citenamefont {Kikkawa}, \citenamefont {Sinova},
  \citenamefont {Nowak}, \citenamefont {Saitoh},\ and\ \citenamefont
  {Kl{\"a}ui}}]{Cra18nac}%
  \BibitemOpen
  \bibfield  {author} {\bibinfo {author} {\bibfnamefont {J.}~\bibnamefont
  {Cramer}}, \bibinfo {author} {\bibfnamefont {F.}~\bibnamefont {Fuhrmann}},
  \bibinfo {author} {\bibfnamefont {U.}~\bibnamefont {Ritzmann}}, \bibinfo
  {author} {\bibfnamefont {V.}~\bibnamefont {Gall}}, \bibinfo {author}
  {\bibfnamefont {T.}~\bibnamefont {Niizeki}}, \bibinfo {author} {\bibfnamefont
  {R.}~\bibnamefont {Ramos}}, \bibinfo {author} {\bibfnamefont
  {Z.}~\bibnamefont {Qiu}}, \bibinfo {author} {\bibfnamefont {D.}~\bibnamefont
  {Hou}}, \bibinfo {author} {\bibfnamefont {T.}~\bibnamefont {Kikkawa}},
  \bibinfo {author} {\bibfnamefont {J.}~\bibnamefont {Sinova}}, \bibinfo
  {author} {\bibfnamefont {U.}~\bibnamefont {Nowak}}, \bibinfo {author}
  {\bibfnamefont {E.}~\bibnamefont {Saitoh}}, \ and\ \bibinfo {author}
  {\bibfnamefont {M.}~\bibnamefont {Kl{\"a}ui}},\ }\href {\doibase
  10.1038/s41467-018-03485-5} {\bibfield  {journal} {\bibinfo  {journal} {Nat.
  Commun.}\ }\textbf {\bibinfo {volume} {9}},\ \bibinfo {pages} {1089}
  (\bibinfo {year} {2018})}\BibitemShut {NoStop}%
\bibitem [{\citenamefont {Sanders}\ and\ \citenamefont
  {Walton}(1977)}]{San77prb}%
  \BibitemOpen
  \bibfield  {author} {\bibinfo {author} {\bibfnamefont {D.~J.}\ \bibnamefont
  {Sanders}}\ and\ \bibinfo {author} {\bibfnamefont {D.}~\bibnamefont
  {Walton}},\ }\href {\doibase 10.1103/PhysRevB.15.1489} {\bibfield  {journal}
  {\bibinfo  {journal} {Phys. Rev. B}\ }\textbf {\bibinfo {volume} {15}},\
  \bibinfo {pages} {1489} (\bibinfo {year} {1977})}\BibitemShut {NoStop}%
\bibitem [{\citenamefont {An}\ \emph {et~al.}(2013)\citenamefont {An},
  \citenamefont {Vasyuchka}, \citenamefont {Uchida}, \citenamefont {Chumak},
  \citenamefont {Yamaguchi}, \citenamefont {Harii}, \citenamefont {Ohe},
  \citenamefont {Jungfleisch}, \citenamefont {Kajiwara}, \citenamefont
  {Adachi}, \citenamefont {Hillebrands}, \citenamefont {Maekawa},\ and\
  \citenamefont {Saitoh}}]{Ant13nam}%
  \BibitemOpen
  \bibfield  {author} {\bibinfo {author} {\bibfnamefont {T.}~\bibnamefont
  {An}}, \bibinfo {author} {\bibfnamefont {V.~I.}\ \bibnamefont {Vasyuchka}},
  \bibinfo {author} {\bibfnamefont {K.}~\bibnamefont {Uchida}}, \bibinfo
  {author} {\bibfnamefont {A.~V.}\ \bibnamefont {Chumak}}, \bibinfo {author}
  {\bibfnamefont {K.}~\bibnamefont {Yamaguchi}}, \bibinfo {author}
  {\bibfnamefont {K.}~\bibnamefont {Harii}}, \bibinfo {author} {\bibfnamefont
  {J.}~\bibnamefont {Ohe}}, \bibinfo {author} {\bibfnamefont {M.~B.}\
  \bibnamefont {Jungfleisch}}, \bibinfo {author} {\bibfnamefont
  {Y.}~\bibnamefont {Kajiwara}}, \bibinfo {author} {\bibfnamefont
  {H.}~\bibnamefont {Adachi}}, \bibinfo {author} {\bibfnamefont
  {B.}~\bibnamefont {Hillebrands}}, \bibinfo {author} {\bibfnamefont
  {S.}~\bibnamefont {Maekawa}}, \ and\ \bibinfo {author} {\bibfnamefont
  {E.}~\bibnamefont {Saitoh}},\ }\href {http://dx.doi.org/10.1038/nmat3628}
  {\bibfield  {journal} {\bibinfo  {journal} {Nat. Mater.}\ }\textbf {\bibinfo
  {volume} {12}},\ \bibinfo {pages} {549} (\bibinfo {year} {2013})}\BibitemShut
  {NoStop}%
\bibitem [{\citenamefont {Serga}\ \emph {et~al.}(2010)\citenamefont {Serga},
  \citenamefont {Chumak},\ and\ \citenamefont {Hillebrands}}]{Ser10jpd}%
  \BibitemOpen
  \bibfield  {author} {\bibinfo {author} {\bibfnamefont {A.~A.}\ \bibnamefont
  {Serga}}, \bibinfo {author} {\bibfnamefont {A.~V.}\ \bibnamefont {Chumak}}, \
  and\ \bibinfo {author} {\bibfnamefont {B.}~\bibnamefont {Hillebrands}},\
  }\href {http://stacks.iop.org/0022-3727/43/i=26/a=264002} {\bibfield
  {journal} {\bibinfo  {journal} {J. Phys. D: Appl. Phys.}\ }\textbf {\bibinfo
  {volume} {43}},\ \bibinfo {pages} {264002} (\bibinfo {year}
  {2010})}\BibitemShut {NoStop}%
\bibitem [{\citenamefont {Bhandari}\ and\ \citenamefont
  {Verma}(1966)}]{Bha66prv}%
  \BibitemOpen
  \bibfield  {author} {\bibinfo {author} {\bibfnamefont {C.~M.}\ \bibnamefont
  {Bhandari}}\ and\ \bibinfo {author} {\bibfnamefont {G.~S.}\ \bibnamefont
  {Verma}},\ }\href {\doibase 10.1103/PhysRev.152.731} {\bibfield  {journal}
  {\bibinfo  {journal} {Phys. Rev.}\ }\textbf {\bibinfo {volume} {152}},\
  \bibinfo {pages} {731} (\bibinfo {year} {1966})}\BibitemShut {NoStop}%
\bibitem [{\citenamefont {Kaganov}\ \emph {et~al.}(1956)\citenamefont
  {Kaganov}, \citenamefont {Lifshitz},\ and\ \citenamefont
  {Tanatarov}}]{Kag56etp}%
  \BibitemOpen
  \bibfield  {author} {\bibinfo {author} {\bibfnamefont {M.}~\bibnamefont
  {Kaganov}}, \bibinfo {author} {\bibfnamefont {I.~M.}\ \bibnamefont
  {Lifshitz}}, \ and\ \bibinfo {author} {\bibfnamefont {L.~V.}\ \bibnamefont
  {Tanatarov}},\ } \href{http://www.jetp.ac.ru/cgi-bin/e/index/e/4/2/p173?a=list}
  {\bibfield  {journal} {\bibinfo  {journal} {J. Exp. Theor. Phys.}\ }\textbf
  {\bibinfo {volume} {31}},\ \bibinfo {pages} {232} (\bibinfo {year}
  {1956})}\BibitemShut {NoStop}%
\bibitem [{\citenamefont {Kabanov}\ and\ \citenamefont
  {Alexandrov}(2008)}]{Kab08prb}%
  \BibitemOpen
  \bibfield  {author} {\bibinfo {author} {\bibfnamefont {V.~V.}\ \bibnamefont
  {Kabanov}}\ and\ \bibinfo {author} {\bibfnamefont {A.~S.}\ \bibnamefont
  {Alexandrov}},\ }\href {\doibase 10.1103/PhysRevB.78.174514} {\bibfield
  {journal} {\bibinfo  {journal} {Phys. Rev. B}\ }\textbf {\bibinfo {volume}
  {78}},\ \bibinfo {pages} {174514} (\bibinfo {year} {2008})}\BibitemShut
  {NoStop}%
\bibitem [{\citenamefont {Xiao}\ \emph {et~al.}(2010)\citenamefont {Xiao},
  \citenamefont {Bauer}, \citenamefont {Uchida}, \citenamefont {Saitoh},\ and\
  \citenamefont {Maekawa}}]{Xia10prb}%
  \BibitemOpen
  \bibfield  {author} {\bibinfo {author} {\bibfnamefont {J.}~\bibnamefont
  {Xiao}}, \bibinfo {author} {\bibfnamefont {G.~E.~W.}\ \bibnamefont {Bauer}},
  \bibinfo {author} {\bibfnamefont {K.-c.}\ \bibnamefont {Uchida}}, \bibinfo
  {author} {\bibfnamefont {E.}~\bibnamefont {Saitoh}}, \ and\ \bibinfo {author}
  {\bibfnamefont {S.}~\bibnamefont {Maekawa}},\ }\href {\doibase
  10.1103/PhysRevB.81.214418} {\bibfield  {journal} {\bibinfo  {journal} {Phys.
  Rev. B}\ }\textbf {\bibinfo {volume} {81}},\ \bibinfo {pages} {214418}
  (\bibinfo {year} {2010})}\BibitemShut {NoStop}%
\bibitem [{\citenamefont {Bezuglyj}\ and\ \citenamefont
  {Shklovskij}(2016)}]{Bez16ltp}%
  \BibitemOpen
  \bibfield  {author} {\bibinfo {author} {\bibfnamefont {A.~I.}\ \bibnamefont
  {Bezuglyj}}\ and\ \bibinfo {author} {\bibfnamefont {V.~A.}\ \bibnamefont
  {Shklovskij}},\ }\href {\doibase 10.1063/1.4962149} {\bibfield  {journal}
  {\bibinfo  {journal} {Low Temp. Phys.}\ }\textbf {\bibinfo {volume} {42}},\
  \bibinfo {pages} {636} (\bibinfo {year} {2016})}\BibitemShut {NoStop}%
\bibitem [{\citenamefont {Liu}\ \emph {et~al.}(2017)\citenamefont {Liu},
  \citenamefont {Xie}, \citenamefont {Yuan},\ and\ \citenamefont
  {Xia}}]{Liu17prb}%
  \BibitemOpen
  \bibfield  {author} {\bibinfo {author} {\bibfnamefont {Y.}~\bibnamefont
  {Liu}}, \bibinfo {author} {\bibfnamefont {L.-S.}\ \bibnamefont {Xie}},
  \bibinfo {author} {\bibfnamefont {Z.}~\bibnamefont {Yuan}}, \ and\ \bibinfo
  {author} {\bibfnamefont {K.}~\bibnamefont {Xia}},\ }\href {\doibase
  10.1103/PhysRevB.96.174416} {\bibfield  {journal} {\bibinfo  {journal} {Phys.
  Rev. B}\ }\textbf {\bibinfo {volume} {96}},\ \bibinfo {pages} {174416}
  (\bibinfo {year} {2017})}\BibitemShut {NoStop}%
\bibitem [{\citenamefont {Bezuglyj}\ and\ \citenamefont
  {Shklovskij}(2018)}]{Bez18pcm}%
  \BibitemOpen
  \bibfield  {author} {\bibinfo {author} {\bibfnamefont {A.~I.}\ \bibnamefont
  {Bezuglyj}}\ and\ \bibinfo {author} {\bibfnamefont {V.~A.}\ \bibnamefont
  {Shklovskij}},\ }\href {http://iopscience.iop.org/10.1088/1361-648X/aacb72}
  {\bibfield  {journal} {\bibinfo  {journal} {J. Phys.: Condens. Matt.}\ }
  (\bibinfo {year} {2018})}\BibitemShut {NoStop}%
\bibitem [{\citenamefont {Anisimov}\ \emph {et~al.}(1970)\citenamefont
  {Anisimov}, \citenamefont {Imas}, \citenamefont {Romanov},\ and\
  \citenamefont {Khodyko}}]{Ani70boo}%
  \BibitemOpen
  \bibfield  {author} {\bibinfo {author} {\bibfnamefont {S.~I.}\ \bibnamefont
  {Anisimov}}, \bibinfo {author} {\bibfnamefont {Y.~A.}\ \bibnamefont {Imas}},
  \bibinfo {author} {\bibfnamefont {G.~S.}\ \bibnamefont {Romanov}}, \ and\
  \bibinfo {author} {\bibfnamefont {Y.~V.}\ \bibnamefont {Khodyko}},\
  }\href@noop {} {\emph {\bibinfo {title} {Action of high-power radiation on
  metals}}}\ (\bibinfo  {publisher} {Moscow, Nauka},\ \bibinfo {year}
  {1970})\BibitemShut {NoStop}%
\bibitem [{\citenamefont {Anisimov}\ \emph
  {et~al.}(1974{\natexlab{a}})\citenamefont {Anisimov}, \citenamefont
  {Kapeliovich},\ and\ \citenamefont {Perelman}}]{Ani74spj}%
  \BibitemOpen
  \bibfield  {author} {\bibinfo {author} {\bibfnamefont {S.~I.}\ \bibnamefont
  {Anisimov}}, \bibinfo {author} {\bibfnamefont {B.~L.}\ \bibnamefont
  {Kapeliovich}}, \ and\ \bibinfo {author} {\bibfnamefont {T.~L.}\ \bibnamefont
  {Perelman}},\ }\href@noop {} {\bibfield  {journal} {\bibinfo  {journal} {Sov.
  Phys. JETP}\ }\textbf {\bibinfo {volume} {39}},\ \bibinfo {pages} {375}
  (\bibinfo {year} {1974}{\natexlab{a}})}\BibitemShut {NoStop}%
\bibitem [{\citenamefont {Fujimoto}\ \emph {et~al.}(1984)\citenamefont
  {Fujimoto}, \citenamefont {Liu}, \citenamefont {Ippen},\ and\ \citenamefont
  {Bloembergen}}]{Fuj84prl}%
  \BibitemOpen
  \bibfield  {author} {\bibinfo {author} {\bibfnamefont {J.~G.}\ \bibnamefont
  {Fujimoto}}, \bibinfo {author} {\bibfnamefont {J.~M.}\ \bibnamefont {Liu}},
  \bibinfo {author} {\bibfnamefont {E.~P.}\ \bibnamefont {Ippen}}, \ and\
  \bibinfo {author} {\bibfnamefont {N.}~\bibnamefont {Bloembergen}},\ }\href
  {\doibase 10.1103/PhysRevLett.53.1837} {\bibfield  {journal} {\bibinfo
  {journal} {Phys. Rev. Lett.}\ }\textbf {\bibinfo {volume} {53}},\ \bibinfo
  {pages} {1837} (\bibinfo {year} {1984})}\BibitemShut {NoStop}%
\bibitem [{\citenamefont {Elsayed-Ali}\ \emph {et~al.}(1987)\citenamefont
  {Elsayed-Ali}, \citenamefont {Norris}, \citenamefont {Pessot},\ and\
  \citenamefont {Mourou}}]{Els87prl}%
  \BibitemOpen
  \bibfield  {author} {\bibinfo {author} {\bibfnamefont {H.~E.}\ \bibnamefont
  {Elsayed-Ali}}, \bibinfo {author} {\bibfnamefont {T.~B.}\ \bibnamefont
  {Norris}}, \bibinfo {author} {\bibfnamefont {M.~A.}\ \bibnamefont {Pessot}},
  \ and\ \bibinfo {author} {\bibfnamefont {G.~A.}\ \bibnamefont {Mourou}},\
  }\href {\doibase 10.1103/PhysRevLett.58.1212} {\bibfield  {journal} {\bibinfo
   {journal} {Phys. Rev. Lett.}\ }\textbf {\bibinfo {volume} {58}},\ \bibinfo
  {pages} {1212} (\bibinfo {year} {1987})}\BibitemShut {NoStop}%
\bibitem [{\citenamefont {Schoenlein}\ \emph {et~al.}(1987)\citenamefont
  {Schoenlein}, \citenamefont {Lin}, \citenamefont {Fujimoto},\ and\
  \citenamefont {Eesley}}]{Sch87prl}%
  \BibitemOpen
  \bibfield  {author} {\bibinfo {author} {\bibfnamefont {R.~W.}\ \bibnamefont
  {Schoenlein}}, \bibinfo {author} {\bibfnamefont {W.~Z.}\ \bibnamefont {Lin}},
  \bibinfo {author} {\bibfnamefont {J.~G.}\ \bibnamefont {Fujimoto}}, \ and\
  \bibinfo {author} {\bibfnamefont {G.~L.}\ \bibnamefont {Eesley}},\ }\href
  {\doibase 10.1103/PhysRevLett.58.1680} {\bibfield  {journal} {\bibinfo
  {journal} {Phys. Rev. Lett.}\ }\textbf {\bibinfo {volume} {58}},\ \bibinfo
  {pages} {1680} (\bibinfo {year} {1987})}\BibitemShut {NoStop}%
\bibitem [{\citenamefont {Akhieser}(1946)}]{Akh46jpu}%
  \BibitemOpen
  \bibfield  {author} {\bibinfo {author} {\bibfnamefont {A.}~\bibnamefont
  {Akhieser}},\ }\href@noop {} {\bibfield  {journal} {\bibinfo  {journal} {J.
  Phys. USSR}\ }\textbf {\bibinfo {volume} {10}},\ \bibinfo {pages} {247}
  (\bibinfo {year} {1946})}\BibitemShut {NoStop}%
\bibitem [{\citenamefont {Jungfleisch}\ \emph {et~al.}(2015)\citenamefont
  {Jungfleisch}, \citenamefont {Chumak}, \citenamefont {Kehlberger},
  \citenamefont {Lauer}, \citenamefont {Kim}, \citenamefont {Onbasli},
  \citenamefont {Ross}, \citenamefont {Kl\"aui},\ and\ \citenamefont
  {Hillebrands}}]{Jun15prb}%
  \BibitemOpen
  \bibfield  {author} {\bibinfo {author} {\bibfnamefont {M.~B.}\ \bibnamefont
  {Jungfleisch}}, \bibinfo {author} {\bibfnamefont {A.~V.}\ \bibnamefont
  {Chumak}}, \bibinfo {author} {\bibfnamefont {A.}~\bibnamefont {Kehlberger}},
  \bibinfo {author} {\bibfnamefont {V.}~\bibnamefont {Lauer}}, \bibinfo
  {author} {\bibfnamefont {D.~H.}\ \bibnamefont {Kim}}, \bibinfo {author}
  {\bibfnamefont {M.~C.}\ \bibnamefont {Onbasli}}, \bibinfo {author}
  {\bibfnamefont {C.~A.}\ \bibnamefont {Ross}}, \bibinfo {author}
  {\bibfnamefont {M.}~\bibnamefont {Kl\"aui}}, \ and\ \bibinfo {author}
  {\bibfnamefont {B.}~\bibnamefont {Hillebrands}},\ }\href {\doibase
  10.1103/PhysRevB.91.134407} {\bibfield  {journal} {\bibinfo  {journal} {Phys.
  Rev. B}\ }\textbf {\bibinfo {volume} {91}},\ \bibinfo {pages} {134407}
  (\bibinfo {year} {2015})}\BibitemShut {NoStop}%
\bibitem [{\citenamefont {Akhiezer}\ \emph {et~al.}(1967)\citenamefont
  {Akhiezer}, \citenamefont {Bar'yakhtar},\ and\ \citenamefont
  {Peletminski}}]{Akh67boo}%
  \BibitemOpen
  \bibfield  {author} {\bibinfo {author} {\bibfnamefont {A.}~\bibnamefont
  {Akhiezer}}, \bibinfo {author} {\bibfnamefont {V.~G.}\ \bibnamefont
  {Bar'yakhtar}}, \ and\ \bibinfo {author} {\bibfnamefont {S.~V.}\ \bibnamefont
  {Peletminski}},\ }\href@noop {} {\emph {\bibinfo {title} {Spin waves}}}\
  (\bibinfo  {publisher} {Moscow, Nauka},\ \bibinfo {year} {1967})\BibitemShut
  {NoStop}%
\bibitem [{\citenamefont {Kapitza}(1941)}]{Kap41jph}%
  \BibitemOpen
  \bibfield  {author} {\bibinfo {author} {\bibfnamefont {P.~L.}\ \bibnamefont
  {Kapitza}},\ }\href@noop {} {\bibfield  {journal} {\bibinfo  {journal} {J.
  Phys.}\ }\textbf {\bibinfo {volume} {4}},\ \bibinfo {pages} {181} (\bibinfo
  {year} {1941})}\BibitemShut {NoStop}%
\bibitem [{\citenamefont {Prudnikov}\ \emph {et~al.}(2002)\citenamefont
  {Prudnikov}, \citenamefont {Brychkov},\ and\ \citenamefont
  {Marichev}}]{Pru02boo}%
  \BibitemOpen
  \bibfield  {author} {\bibinfo {author} {\bibfnamefont {A.~P.}\ \bibnamefont
  {Prudnikov}}, \bibinfo {author} {\bibfnamefont {Y.~A.}\ \bibnamefont
  {Brychkov}}, \ and\ \bibinfo {author} {\bibfnamefont {O.~I.}\ \bibnamefont
  {Marichev}},\ }\href@noop {} {\emph {\bibinfo {title} {Integrals and series.
  Elementary functions}}}\ (\bibinfo  {publisher} {Moscow, Nauka},\ \bibinfo
  {year} {2002})\BibitemShut {NoStop}%
\end{thebibliography}
%

\end{document}